\documentclass[12pt]{iopart}

\usepackage{iopams,setstack}   
\usepackage{amsthm,bm}

\newcommand{\mi}{\rmi}
\newcommand{\me}{\rme}

\newcommand{\R}{\mathbb{R}}
\newcommand{\Hil}{\mathcal{H}}
\newcommand{\ket}[1]{| #1 \rangle}
\newcommand{\Bigket}[1]{\Bigl| #1 \Bigr\rangle}
\newcommand{\bra}[1]{\langle #1 |}
\newcommand{\Bigbra}[1]{\Bigl\langle #1 \Bigr|}
\newcommand{\Op}[1]{\hat{#1}}
\newcommand{\Id}{\mathbf 1}
\newcommand{\OpU}{\Op{\mathcal U}}
\newcommand{\OpS}{\Op{\mathcal S}} 
\newcommand{\fe}[1]{\underline{#1}}
\newcommand{\braket}[2]{\langle #1 | #2 \rangle}
\newcommand{\Av}[1]{\langle #1 \rangle}
\newcommand{\vac}{\ket{\emptyset}}
\newcommand{\per}{\mathop\mathrm{per}} %\DeclareMathOperator{\per}{per}
\newcommand{\PSym}{\mathop{\mathcal P}} %\DeclareMathOperator{\PSym}{\mathcal P}

\begin{document}

\review{Permanents, Bosons and Linear Optics}

\author{Alexander Yu.\ Vlasov}%
\address{Federal Radiology Center (IRH)\\
	197101, Mira Street 8, Saint Petersburg, Russia}%
\ead{\mailto{qubeat@mail.ru}}%

%\date{}

\sloppy

\begin{abstract}
Particular complexity of linear quantum optical networks is deserved recently 
certain attention due to possible implications for theory of quantum computation.
Two relevant models of bosons are discussed in presented work.
Symmetric product of Hilbert spaces produces rather abstract model. 
The second one is obtained by quantization of harmonic oscillator. 
In contrast to considered bosonic processes, so-called ``fermionic linear optics'' is 
effectively simulated on classical computer. The comparison
of bosonic and fermionic case clarifies the controversy
and the more elaborated oscillator model provides a deeper analogy.
\end{abstract}

\vspace{2pc}

%\maketitle

%\ioptwocol

\section{Introduction}
\label{Sec:Intro}

{\em Boson sampling} \cite{AA} is a formal model of nonuniversal quantum computations with 
noninteracting bosons suggested as a system that cannot be 
effectively simulated by classical computers. 
The model is much simpler than scalable universal quantum computer, but it might provide
a principle evidence of so-called {\em quantum supremacy} important for theory of complexity.

This paper is devoted to two issues. The first one
is existence of a couple alternative quantum models 
relevant to description of the {\em boson sampling}. 
The second question is comparison with ``fermionic'' quantum circuits effectively 
simulated by classical computer.

In \sref{Sec:Naive} the {\em ``abstract'' bosons model} is briefly
discussed and after short reminder about linear optics in \sref{Sec:Lino} 
more elaborated {\em oscillator model} is discussed in \sref{Sec:Osc}.
The comparison of ``bosonic'' and ``fermionic'' case is provided
in \sref{Sec:CompFerm}.

\section{Abstract bosons model}
\label{Sec:Naive}

An abstract quantum system with $d$ states is described by 
$d$-dimensional Hilbert space $\Hil$. 
An arbitrary state can be expressed 
using Dirac notation \cite{Dirac} with a basis
$\ket{1},\ket{2}, \dots,\ket{d}$
\begin{equation}
\ket{\psi} = \sum_{k=1}^d c_k \ket{k},
\quad \sum_{k=1}^d |c_k|^2 = 1.
\label{dudit}
\end{equation}

Group U$(d)$ of unitary matrices describes an evolution of the state. 
In such description a global phase does not matter \cite{Dirac}
and the group SU$(d)$ of unitary matrices with unit determinant
may be used instead of U$(d)$ 
\begin{equation}
\ket{\psi'} = \Op{u} \ket{\psi}, \quad
c'_j = \sum_{k=1}^d u_{jk} c_k,
\quad \Op{u} \in \mathrm{SU}(d).
\label{upsi}
\end{equation}

Let us denote $S^n(\Hil)$ linear space of symmetric $n$-tensors, 
$\dim S^n(\Hil) = C^{d+n-1}_n$. Symmetric product of tensors also 
can be used with similar purpose. The method is related to
linear space of polynomials of degree $n$ with $d$ variables 
\cite{KM}. For $n=1$ such a space is simply $\Hil$.
The $S^n(\Hil)$ can be considered as a state of system with $n$ 
indistinguishable bosons \cite{KM}. 
An element of $S^n(\Hil)$ represented as symmetric product of $n$ vectors 
from $\Hil$ is called {\em decomposable}. It may be treated as 
a system with $n$ {\em noninteracting} bosons.

Two methods of indexing are used for basis of $S^n(\Hil)$ \cite{Weyl}. 
The first one is sequence $\bm j$ with $n$ indexes
\begin{equation}
{\bm j} = (j_1,\ldots,j_n), 
\quad 1 \leq j_1 \leq j_2 \leq \cdots \leq j_n \leq d.
\label{jind}  
\end{equation}
Such notation is obtained from initial basis of tensor product after symmetrization
of indexes. It may be rewritten due to polynomial representation \cite{Weyl}
\begin{equation}
 x_{j_1}x_{j_2} \ldots x_{j_n} = x_1^{r_1^{\bm j}} \ldots x_d^{r_d^{\bm j}}, 
 ~ r_1^{\bm j}+r_2^{\bm j}+\cdots+r_d^{\bm j} = n, 
 %\quad r_k^{\bm j} \ge 0,
 \label{dind}
\end{equation}
where $r_k^{\bm j} \ge 0$ is number of indexes $k$ in the sequence $\bm j$.
The numbers $r_k^{\bm j}$ produce an alternative representation as a vector 
$\bm{r^j}$ with $d$ elements.
The notation is convenient for expression of standard normalization 
for basic vectors in space of symmetric tensors or polynomials
\begin{equation}
x^S_{\bm j} = \frac{x_1^{r_1^{\bm j}} \ldots x_d^{r_d^{\bm j}}}{\sqrt{\Gamma_{\bm j}}},
% \longrightarrow\ket{r_1^{\bm j},\ldots,r_d^{\bm j}}, 
 \quad
 \Gamma_{\bm j} = r_1^{\bm j}!\ldots r_d^{\bm j}!
\label{symnorm} 
\end{equation}
and the element of $S^n(\Hil)$ may be considered as a complex vector
with $C^{d+n-1}_n$ components in a basis $\ket{x^S_{\bm j}}$.
Action of unitary group on each $\Hil$ for such normalization corresponds to
an unitary representation on the $S^n(\Hil)$ with respect to usual
Hermitian scalar product and the unitarity \cite{Weyl} is natural due to   
\begin{equation}
\Bigl(\sum_{k=1}^d x_k\bar{x}_k\Bigr)^n 
= \!\!\!\sum_{\sum r_k = n}\!\!\!{\frac{n!\prod x_k^{r_k} \bar{x}_k^{r_k}}{r_1!\ldots r_d!}}
=n!\sum_{\bm j \in S}{x^S_{\bm j}\bar{x}^S_{\bm j}}.
\label{normpow}
\end{equation}

For already mentioned earlier {\em decomposable} elements of $S^n(\Hil)$ relations 
between the scalar products on $S^n(\Hil)$ and $\Hil$ may be expressed using permanents \cite{Perm}.
Let us recall, that the {\em permanent} of $n \times n$ matrix $A$ is 
\begin{equation}
 \per(A) = \sum_{\sigma}\prod_{j=1}^n A_{j,\sigma(j)},
\label{perm} 
\end{equation} 
where $\sigma$ denotes all possible permutations of indexes. The only difference with 
determinant is lack of minus signs for some terms. However, unlike the determinant the permanent
is an example of computationally hard problem \cite{ValPer}.

Transformation of product \eref{dind} for change of variables
$x'_j = \sum_{k=1}^d M_{jk}x_k$ can be expressed 
with some polynomials $\PSym(M)$ of degree $n$ with elements of the matrix $M$
\begin{equation}
x'_{j_1}x'_{j_2} \ldots x'_{j_n} 
= \!\sum_{k_1,\ldots,k_n}\! \PSym(M)^{k_1,\ldots,k_n}_{j_1,\ldots,j_n} x_{k_1}x_{k_2} \ldots x_{k_n}.
\label{prpoly}
\end{equation}
A simple case suitable for further examples is
$n=d$ with consequent indexes ${\bm k} = {\bm j} = (1,2,\ldots,n)$, 
when the polynomial simply equal to permanent
\begin{equation}
\PSym(M)^{1,\ldots,n}_{1,\ldots,n} = \per(M).
\label{simperm}
\end{equation}
More general case should be discussed elsewhere and analogues of \eref{simperm} also 
could be expressed using permanents of matrices composed from rows and columns of $M$.

An application of such an abstract model for discussion about permanent and determinant
complexity may be found in \cite{TT96} together with suggestion about 
ineffectiveness of finding permanents using quantum processes with bosons because of
big variance of measurement outcomes.

\section{Linear optics}
\label{Sec:Lino}

{\em The abstract bosons model} described above says a little about physical
realization. Alternative way of producing expressions related with
permanent may be obtained using quantum model of linear optical networks 
\cite{Sch04}. It may have relevance with theory of quantum 
computing \cite{loqc} and computational 
complexity of linear optics was discussed further coining the term 
{\em boson sampling} \cite{AA} stimulating 
the series of experiments \cite{Sci1,Sci2,NatP1,NatP2}.

An evident distinction in formulation of such a model in comparison with
{\em the abstract bosons model} from \sref{Sec:Naive} is definition
of some basic transformations using {\em creation and annihilation operators} 
$\Op{a}_j$, $\Op{a}_j^\dag$, $j=1,\ldots,d$ \cite{Sch04,loqc}.

The approach is quite natural, because with such a notation linear optical
network with conserved total photon number corresponds to transformation \cite{loqc}
\begin{equation}
\Op{a}'_j = \sum_{k=1}^d U_{jk} \Op{a}_k,
\quad U \in \mathrm U(d),
\quad j,k = 1,\ldots,d.
\label{Ua} 
\end{equation}

Due to some analogy between \eref{Ua} and \eref{upsi} they might
cause similarity in formal calculations, but meanings of the expressions
are quite different. The \eref{Ua} describes transformation of operators
in Heisenberg representation, but \eref{upsi} is applied to states.

Transformation of some operator $\Op{A}$ due to \eref{upsi} in the {\em abstract bosons model}
from \sref{Sec:Naive} in {\em Heisenberg representation} \cite{BSQF} would be 
\begin{equation}
\Op{A}' = \Op{u} \Op{A} \Op{u}^\dag.
\label{uAu}
\end{equation}
Here $\Op{u}$ produces the same result as $\me^{\mi\phi}\Op{u}$ reaffirming 
sufficiency of using $\Op{u} \in \mathrm{SU}(d)$.
In \eref{Ua} such compensation of phases is dispensable and so
the whole unitary group U$(d)$ may be implemented, but it is rather
a hint on more essential difference between the models discussed further.

In fact, a model with symmetrization of states very
similar with {\em abstract bosons model} from \sref{Sec:Naive} with
infinite-dimensional space $\Hil$ sometimes used in quantum optics 
as well \cite{SZ97}.
It can be asked in turn, how to rewrite \eref{Ua} in a way
similar with \eref{uAu}. 
For such a purpose in next section transformation with conservation of total 
photon number is considered as particular case of most general linear Bogoliubov
transformations \cite{loqc,BSQF} {\em without} such requirement.

\section{Oscillator model}
\label{Sec:Osc}

\subsection{Schr\"odinger description}

The model of quantum harmonic oscillator is recollected below with
a brief excursus into theory of symplectic and metaplectic groups 
necessary for applications to linear optics \cite{ADMS,GSsym,Foll89}.
Let us consider {\em infinite-dimensional} Hilbert space $\Hil$ and 
operators $\Op{q}$, $\Op{p}$ of coordinate and momentum. 
In Schr\"odinger description the $\Hil$ is associated with
space of {\em wave functions} $\psi(q) \in L^2(\R)$ (square integrable)
and the operators $\Op{q}$, $\Op{p}$ are defined as
\begin{equation}
 \Op{q} \colon \psi(q) \to q \,\psi(q), \qquad
 \Op{p} \colon \psi(q) \to -\mi \partial\psi(q)/\partial q
\label{pqsch} 
\end{equation}
with canonical commutation relation (CCR)
\begin{equation}
[\Op{p},\Op{q}] = \Op{p}\Op{q} - \Op{q}\Op{p} = -\mi,
\label{CCR}
\end{equation}
where system of units with $\hbar = 1$ is used for simplicity. 
The generalization on set of operators
$\Op{q}_k$, $\Op{p}_k$, $k=1,\ldots,d$ with CCR
\begin{equation}
[\Op{p}_j,\Op{q}_k] = -\mi \delta_{jk},
\quad j,k = 1,\ldots,d
\label{CCRs}
\end{equation}
is straightforward
using space $L^2(\R^d)$ of wave functions with $d$ variables.

\subsection{Symplectic group}
\label{Sec:Sympl}

A real $2d \times 2d$ matrix $A$ preserving bilinear form 
\begin{equation}
(x,y)_\Lambda = \sum_{k=1}^d (x_k y_{d+k} - x_{d+k} y_k)
\label{skew}
\end{equation}
%for vectors $ x,y \in \R^{2d}$ 
is called {\em symplectic}.
It also has property \cite{Post}
\begin{equation}
	A^T J A = J,
\label{AJA}	
\end{equation}
where $A^T$ is transposed matrix and $J$ is $2d \times 2d$ matrix
\begin{equation}
 J = \left(\begin{array}{cc}
   ~\mathbf 0_d & \Id_d \\ -\Id_d & \mathbf 0_d
    \end{array}\right)
\label{matJ} 
\end{equation}
with $\mathbf 0_d$ and $\Id_d$ are zero and unit $d \times d$ matrices.
A composition of such matrices also satisfies \eref{AJA} and, so, the 
{\em symplectic group} Sp$(2d,\R)$ 
\cite{ADMS,GSsym} is defined\footnote{Slightly different parametrization Sp$(d,\R)$ 
	is used in \cite{Foll89,Post}.} in such a way.

Due to an analogy with orthogonal matrices $R \in \mathrm{SO}(d)$ satisfying $R^T R = \mathbf 1$ and
preserving Euclidean norm $(x,y)_E=\sum x_k y_k$ a matrix with property \eref{AJA} and arbitrary
$J$ sometimes is called {\em J-orthogonal}. Such a definition includes both orthogonal (for unit $J$)
and symplectic matrices \cite{Post}.

Matrices preserving both symplectic \eref{skew} and Euclidean forms belong to
{\em orthogonal symplectic group} defined as intersection
 Sp$(2d,\R) \cap \mathrm{SO}(2d)$. The group is isomorphic
with unitary group U$(d)$ \cite{Post}. Indeed, if to
consider complex variables 
\begin{equation}
	z_k = x_k +\mi x_{k+d}, \quad \bar{z}_k = x_k - \mi x_{k+d},
%	\quad k = 1,\ldots,d
\label{x2z}	
\end{equation}
the real and imaginary
part of Hermitian complex scalar product correspond 
to Euclidean and symplectic forms, respectively 
$\braket{z}{z'} = (x,x')_E+\mi(x,x')_\Lambda$.

\smallskip

To show relation of symplectic group to CCR let us write $\Op{q}_k$, $\Op{p}_k$
as formal vector of operators with $2d$ elements \cite{ADMS}
\begin{equation}
 \Op{w} = (\Op{w}_k) = (\Op{q}_1, \ldots, \Op{q}_d,\Op{p}_1, \ldots, \Op{p}_d).
\label{w2d} 
\end{equation}
\Eref{CCRs} can be rewritten in such a case as
\begin{equation}
[\Op{w}_j,\Op{w}_k] = -\mi J_{jk},
\quad j,k = 1,\ldots,2d,
\label{sCCR}
\end{equation}
where $J_{jk}$ are elements of matrix $J$ \eref{matJ}.
Due to such property $2d$ operators $\Op{w}'_j$
\begin{equation}
\Op{w}'_j = \sum_{k=1}^{2d} S_{jk} \Op{w}_k,
\quad j,k = 1,\ldots,2d,
\label{SymSum}
\end{equation} 
also satisfy \eref{sCCR} if matrix $S \in \mathrm{Sp}(2d,\R)$.

\subsection{Metaplectic group}

Both the sets of operators $\Op{w}_j$ and $\Op{w}'_j$ related by \eref{SymSum} satisfy
some form of CCR \eref{sCCR} and in agreement with general results
about uniqueness of CCR they should be {\em unitary equivalent}, 
{\em i.e.}, for any matrix $S$ in \eref{SymSum} some unitary operator $\OpU_S$ 
should provide transformation \cite{ADMS,GSsym,Foll89}
\begin{equation}
  \Op{w}'_j = \OpU_S \Op{w}_j \OpU^{-1}_S,
% \quad \OpU_S \OpU^\dag_S = \Id,
  \quad j = 1,\ldots,2d.
\label{MetaU}  
\end{equation}
Due to \eref{MetaU} $\OpU_S$ and $\me^{\mi \phi}\OpU_S$ correspond to
the same matrix $S$, but such a phase freedom may be withdrawn
and the only inevitable ambiguity is a sign $\pm\OpU_S$. The group 
producing such a 2--1 homomorphism on Sp$(2d,\R)$ is known as {\em metaplectic}, 
Mp$(2d,\R)$ \cite{ADMS,GSsym,Foll89}. 
The unitary representation of Mp$(2d,\R)$ used in \eref{MetaU} is not 
a {\em finite-dimensional} matrix group, but can be expressed 
%in Schr\"odinger description with infinite-dimensional Hilbert space 
by exponents with appropriate linear combinations of $\Op{w}_j\Op{w}_k$.

The symplectic group is relevant also to classical optics, but
metaplectic group is essential in quantum case 
\cite{ADMS,GSsym,Foll89}. The group Mp$(2d,\R)$ {\em describes most general
linear optical networks with $d$ modes} and a subgroup of transformations
with conservation of total photon number is discussed below.

\subsection{Annihilation and creation operators}

The transformations respecting also Euclidean norm, {\em i.e.},
sum of Hamiltonians of harmonic oscillators
\begin{equation}
 \Op{H} = \frac{1}{2}\sum_{k=1}^{2d}\Op{w}^2_k = \sum_{k=1}^d\Op{H}_k,\quad
 \Op{H}_k = \frac{1}{2}(\Op{p}_k^2 + \Op{q}_k^2) 
\label{Hsum} 
\end{equation}
correspond to already mentioned {\em orthogonal symplectic group}
isomorphic with unitary group U$(d)$.
Complex coordinates \eref{x2z} now associated with {\em annihilation and
creation (``ladder'') operators}
\begin{equation}
 \Op{a}_k = \frac{1}{\sqrt{2}}(\Op{q}_k + \mi \Op{p}_k), \quad
 \Op{a}_k^\dag = \frac{1}{\sqrt{2}}(\Op{q}_k - \mi \Op{p}_k).
%  \quad k = 1,\ldots,d.
\label{lad}  
\end{equation}

Equations~(\ref{Hsum}, \ref{lad}) are widely accepted \cite{ADMS,GSsym,Foll89}, 
consistent with rather standard map to U$(d)$ \cite{Post}
and used further in this work for simplicity
instead of more general versions also relevant to quantum optics \cite{BSQF,LLQED}
\begin{equation}
\eqalign{\Op{H} = \frac{1}{2}(\Op{p}^2 + \omega^2\Op{q}^2), \cr \qquad 
\Op{a} = \frac{1}{\sqrt{2\omega}}(\omega\Op{q} + \mi \Op{p}),\quad
\Op{a}^\dag = \frac{1}{\sqrt{2\omega}}(\omega\Op{q} - \mi \Op{p}).}
\label{om} 
\end{equation}

\medskip

The oscillator model relates group U$(d)$ treated as a subgroup of Sp$(2d,\R)$
with conservation of ``total photon number'' defined by operator 
\begin{equation}
 \Op{N} = \sum_{j=1}^d\Op{N}_j, \quad
 \Op{N}_j = \Op{a}_j^\dag\Op{a}_j.
\label{PhNum} 
\end{equation}

Instead of \eref{SymSum} already mentioned earlier should be used \eref{Ua} 
and expression for $\Op{a}_j^\dag$ is obtained using Hermitian conjugation
\begin{equation}
\Op{a}'_j = \sum_{k=1}^d U_{jk} \Op{a}_k,\quad
\Op{a}^{\prime\dag}_j = \sum_{k=1}^d U^*_{jk} \Op{a}^\dag_k,\quad
U \in \mathrm U(d).
%\quad  j,k = 1,\ldots,d.
\label{USum}
\end{equation}

\Eref{SymSum} rewritten in such a way without requirement about photon number
conservation would include terms $\Op{a}^\dag$, $\Op{a}$ in both sums \eref{USum} 
reproducing most general Bogoliubov transformations.

In such an approach subgroup of Mp$(2d,\R)$ corresponding to U$(d)$ 
sometimes is denoted as MU$(d)$ \cite{Noncom}
and can be expressed by exponents with linear combinations of $\Op{a}_j^\dag\Op{a}_k$.
An analogue of \eref{MetaU} is
\begin{equation}
\Op{a}'_j = \OpU_U \Op{a}_j \OpU^{-1}_U,
\quad j = 1,\ldots,d,
\quad \OpU \in \mathrm{MU}(d).
\label{MUU}  
\end{equation}
with conjugated expression for $\Op{a}_j^\dag$.

The MU$(d)$ is double cover of U$(d)$ with sign ambiguity
inherited from relation between Mp$(2d,\R)$ and Sp$(2d,\R)$ \cite{Foll89,Noncom}.
The additional subtle problems may appear, because using some formal manipulations
with $\OpU_U$ an one-to-one map with U$(d)$ could be obtained, but
it cannot be extended to the whole Sp$(2d,\R)$ \cite{Foll89}. 

The difficulties with unitarity after application
of discussed trick to get rid of double cover would produce
rather unrealistic model of photons with prohibition to use squeezing 
transformations. Even if requirement of particle
number conservation is justified for model with a massive bosons the approach
discussed in this section has other important differences from 
the {\em abstract bosons model} introduced in \sref{Sec:Naive}.

A noticeable formal distinction is an additional phase parameter, because action
of U$(d)$ is not reduced to SU$(d)$. The nontrivial structure
of $\OpU_U$ is manifested here, because \eref{MUU} is not sensitive to formal
phase multiplier $\me^{\mi\phi}\OpU_U$. So action of phase multiplier
on matrix $U$ in \eref{USum} is implemented in alternative way and
``embedded'' directly into structure of $\OpU$ via additional term 
proportional to 
%Hamiltonian $\Op{H}$ \eref{Hsum} or 
total photon number operator $\Op{N}$ \eref{PhNum} in the exponent with quadratic 
expressions for elements of group mentioned earlier.

Yet another important property of used model with double cover and sign 
ambiguity due to \eref{MetaU}, \eref{MUU} is similarity with relation 
between orthogonal and Spin groups \cite{Foll89,Noncom}.

\section{Comparison with formal fermionic model}
\label{Sec:CompFerm}

The theory of Spin groups is relevant with
question about difference in complexity between bosons and 
fermionic model often associated with {\em matchgate circuits} effectively simulated on
classical computer.
The matchgate model was introduced in \cite{Val1} with reformulation
using formal fermionic model in \cite{TD2,Kni1}. Similar fermionic model
and equivalent approach with Spin group and Clifford algebras
was also applied earlier to quantum computing problems in 
\cite{BK0,Kit0,Vla99,Vla0}. 

The theme was further developed in works devoted to effective classical 
simulation of such a class of quantum circuits and they are
often described using {\em fermionic annihilation and creation operators} 
\cite{DT4,Joz8,Joz9}. 
The notation $\Op{\fe{a}}_j$, $\Op{\fe{a}}_j^\dag$ is used here for such operators
for distinction from bosonic case.
The analogues of $\Op{w}_j$ \eref{w2d} here are $2d$ {\em generators of Clifford 
algebra}
\begin{equation}
 \Op{c}_{2j-1} = \Op{\fe{a}}_j + \Op{\fe{a}}_j^\dag, \quad
 \Op{c}_{2j} = -\mi(\Op{\fe{a}}_j - \Op{\fe{a}}_j^\dag).
\label{clif} 
\end{equation}
Transformation properties operators \eref{clif} are similar with \eref{SymSum}, \eref{MetaU} 
%or \eref{USum} with \eref{MUU} in bosonic case
and may be written as
\begin{equation}
 \OpS_R \Op{c}_j \OpS_R^{-1} =
 \sum_{k=1}^{2d} R_{jk} \Op{c}_k,
\label{SpinR}  
\end{equation}
but $R_{jk}$ now are elements of orthogonal matrix and $\OpS_R \in \mathrm{Spin}(2d)$
is corresponding to matchgates (or ``fermionic'') quantum 
circuit. The $\OpS_R$ has matrix representation, but number of
components is exponentially bigger than in matrix $R$, because
it corresponds to quantum network with $d$ qubits.

A restricted case of transformation conserving number of fermions
discussed in \cite{TD2} provides even closer analogy with boson case. 
If the evolution is expressed as exponent of Hamiltonian
with linear combination of $\Op{\fe{a}}_j^\dag \Op{\fe{a}}_k$
then the fermionic operators are transformed \cite{TD2}
by direct analogue of \eref{USum}
\begin{equation}
\OpS_U \Op{\fe{a}}_j^\dag \OpS_U^{-1} =
\sum_{k=1}^d U_{jk} \Op{\fe{a}}_k^\dag,
\label{USpin}  
\end{equation}
with $\OpS_U$ is from subgroup of Spin$(2d)$ corresponding to
restriction of SO$(2d)$ on ``realification'' of SU$(d)$, 
{\em i.e.}, orthogonal symplectic group mentioned earlier.

\medskip
  
Two cases should be taken into account for comparison of complexity fermionic and bosonic models.
The {\em single-mode measurement} in terminology of
\cite{DT4} is the first one and mainly used in \cite{Joz8,Joz9,Vla3n}.
In the concise form the setup \cite{Joz9} for such 
experiment is {\em any computational basis 
state $\ket{x_1 \ldots x_d}$ as the input} and {\em a final measurement of
an arbitrary qubit in the computational basis as the output}.
For arbitrary state $\ket{\psi}$ such a measurement may be described
by probability $p_1^{(k)}$ to obtain {\sf 1} as a result of measurement for 
a qubit with index $k$ 
\begin{equation}
 p_1^{(k)} = \bra{\psi}\Op{\fe{a}}_k^\dag\Op{\fe{a}}_k\ket{\psi}
 % = \bra{\psi}\Op{\fe{N}}_k\ket{\psi}
 \label{prob1}
\end{equation}
and an analogue of such expression in bosonic case  
\begin{equation}
 \bra{\Psi}\Op{a}_k^\dag\Op{a}_k\ket{\Psi} =
 \bra{\Psi}\Op{N}_k\ket{\Psi} = \Av{N_k}
\label{AvNum}
\end{equation}
is corresponding to expectation value for {\em number of particles} in
a mode $k$. 
After application of linear optical network 
new expectation values $\Av{N'_k}$ are
\begin{equation}
\eqalign{\ket{\Psi'} = \OpU\ket{\Psi},  \quad
\Av{N'_k} &=\bra{\Psi}\OpU^\dag\Op{a}_k^\dag\Op{a}_k\OpU\ket{\Psi} \cr
&= \bra{\Psi}\OpU^{-1}\Op{a}_k^\dag\OpU\,\OpU^{-1}\Op{a}_k\OpU\ket{\Psi}.}
\label{MUPsiN}
\end{equation}
It can be rewritten using \eref{USum} and \eref{MUU}
with matrix $U$ corresponding to operator $\OpU$
\begin{equation}
\eqalign{\Av{N'_k} &= \Bigbra{\Psi}\Bigl(\sum_{j=1}^d U_{kj}\Op{a}_j^\dag\Bigr)
 \Bigl(\sum_{j=1}^d U^*_{kj}\Op{a}_j\Bigr)\Bigket{\Psi} \cr
  &= \sum_{j,j'=1}^d U_{k,j}U_{k,j'}^*\bra{\Psi}\Op{a}_j^\dag\Op{a}_{j'}\ket{\Psi}.}
\label{UPsiN}
\end{equation} 
 If an effective way to calculate
 $\bra{\Psi}\Op{a}_j^\dag\Op{a}_{j'}\ket{\Psi}$ exists for given input state
 $\ket{\Psi}$ then the $\Av{N'_k}$ also can be calculated in poly time
 using \eref{UPsiN} and the methods are similar with  matchgate model 
 \cite{Joz8,Joz9}. 

The {\em Fock states} \cite{SZ97} can be considered as an alternative for 
{\em computational basis} for input state for a case with $n$ bosons 
and $d$ modes  
\begin{equation}
 \ket{\Psi_{\bm j}} = 
 \frac{\Op{a}_{j_1}^\dag\Op{a}_{j_2}^\dag \cdots \Op{a}_{j_n}^\dag}{\sqrt{\Gamma_{\bm j}}}\vac,
 \quad 1 \leq j_1  \cdots \leq j_n \leq d 
\label{FockSt}  
\end{equation}
with $\vac$ is vacuum state.
The coefficients $\sqrt{\Gamma_{\bm j}}$ can be derived from expressions for ladder operators
\cite{SZ97} and {\em coincide with normalization} \eref{symnorm} 
for vector $\bm{r^j} = (r_1^{\bm j}, \ldots,r_d^{\bm j})$ defined by \eref{dind} 
for given sequence ${\bm j} = (j_1,j_2,\ldots,j_n)$ introduced in \eref{jind}. 
So, finally 
\begin{equation}
\ket{\Psi_{\bm j}} = 
\frac{(\Op{a}_1^\dag)^{r_1^{\bm j}}(\Op{a}_2^\dag)^{r_2^{\bm j}} \cdots %
(\Op{a}_d^\dag)^{r_d^{\bm j}}\,\vac}{\sqrt{r_1^{\bm j}!r_2^{\bm j}!\ldots r_d^{\bm j}!}}.
\label{FockStRep}  
\end{equation}
For such states  $\bra{\Psi_{\bm j}}\Op{a}_k^\dag\Op{a}_{k'}\ket{\Psi_{\bm j}}$
may be effectively calculated.

\smallskip

The other case is {\em multi-mode} measurement \cite{TD2,DT4}. The analogues
of such model in bosonic case would use instead of \eref{AvNum} polynomials with
$\Op{a}_k^\dag$, $\Op{a}_k$ of higher degree. So, different computational complexity 
of determinant and permanent may be really essential in some examples.

It may be written $\OpU\vac = \theta_{\OpU}\vac$ with some phase multiplier $|\theta_{\OpU}|=1$,
because $\OpU$ is expressed as an exponent with linear combinations of $\Op{a}_j^\dag\Op{a}_k$
and for vacuum state $\Op{a}_j\vac = 0$.
Therefore, 
\begin{equation}
\eqalign{
	\OpU\ket{\Psi_{\bm j}} &=
\frac{1}{\sqrt\Gamma_{\bm j}}\,\OpU\Op{a}_{j_1}^\dag\Op{a}_{j_2}^\dag \cdots \Op{a}_{j_n}^\dag\OpU^{-1}
 \OpU\vac \cr
&=\frac{\theta_{\OpU}}{\sqrt\Gamma_{\bm j}}\,\OpU\Op{a}_{j_1}^\dag\Op{a}_{j_2}^\dag \cdots \Op{a}_{j_n}^\dag\OpU^{-1}\vac.}
 \label{OUFock} 
\end{equation}
After further application to each multiplier $\OpU\Op{a}_{j_k}^\dag\OpU^{-1}$ 
and expansion as sums using \eref{USum} expressions with permanents may be obtained.

Anyway, \eref{AvNum} and analogues with small number of ladder operators
may hide complexity arising from application of 
linear optical network to Fock states.
An example is \eref{MUPsiN} or \eref{UPsiN} that may be used for 
effective classical computation of single-mode measurement outcomes 
(expected average number of photons $\Av{N'_k}$ for each mode $k$) after 
application of linear optical networks to Fock states.
  
The permanent complexity is more essential in expressions for ``transition'' amplitudes
(and probabilities) between two Fock states 
\begin{equation}
\alpha_{\bm{kj}} = \bra{\Psi_{\bm k}}\OpU\ket{\Psi_{\bm j}},
%= \frac{\theta_{\OpU}\per(U^{\bm k, \bm j})}{\sqrt{\Gamma_{\bm k}}\sqrt{\Gamma_{\bm j}}\varpi_{\bm k,\bm j}},
\quad p_{\bm{kj}} = |\alpha_{\bm{kj}}|^2. 
%= \frac{{|\per(U^{\bm k, \bm j})|}^2}{\Gamma_{\bm k}\Gamma_{\bm j}\varpi_{\bm k,\bm j}^2}.
\label{trans}
\end{equation}

Basic example with consequent indexes in $\bm k$ and $\bm j$ corresponds to \eref{simperm}.
%with matrix $U^{\bm k, \bm j}$ is defined by \eref{polyperm}.
For calculation of probabilities $p_{\bm{kj}}$ a phase $|\theta_{\OpU}|=1$ from
\eref{OUFock} should be omitted. An alternative consideration with 
commutative polynomials resembling an {\em abstract boson model} 
discussed earlier in \sref{Sec:Naive} may be found in \cite{AA,AAR11}. 

\smallskip

Analogue models with preserving number of fermions was studied in \cite{TD2}. 
The same $d$-dimensional unitary group \eref{USpin} again can handle the evolution, but 
determinants are used instead of permanents and expressions for fermionic amplitudes 
may be efficiently evaluated \cite{TD2}. The analogue of \eref{trans} in fermionic case
directly coincides with determinant.  So, for fermions {\em multi-mode measurement} 
amplitudes also can be efficiently evaluated due to absence of problems with permanent 
calculation discussed earlier for bosons.

\section{Conclusion}

Some topics relevant to consideration of complexity of simulation of quantum
processes with boson were discussed in this paper. A distinction 
between an {\em abstract bosons model} and
more elaborated approach with {\em quantum harmonic oscillator} was emphasized. 
The treatment of real photonic system may require consideration of even subtler problems
that should be discussed elsewhere.

% The consideration of a few different models produced wider view on the topic.
Let us recollect similar equations for transformations of some sets of states
$\ket{\psi_k}$ and  operators $\Op{\omega}_j$ relevant to many examples 
discussed above using notation 
\begin{equation}
	\ket{\psi'_k} = \OpS \ket{\psi_k}, \quad
	\Op{\omega}'_j = \OpS \Op{\omega}_j \OpS^{-1} = \sum_k S_{jk}\Op{\omega}_k.
\label{GenTrans}	
\end{equation}
For oscillator model of general linear quantum optical network
operators $\Op{\omega}_j = \Op{w}_j$ are defined in \eref{w2d} and the $S_{jk}$ is symplectic
matrix \eref{SymSum} with $\OpS$ is unitary operator $\OpU_S$ \eref{MetaU} from metaplectic group.
For an important case with conserving number of bosons $\Op{\omega}_j$ correspond to
annihilation and creation operators, $\OpS$ was denoted as $\OpU_U$ in \eref{MUU}, and
$S_{jk}$ is unitary matrix $U_{jk}$ in \eref{USum}.

In \sref{Sec:CompFerm} about fermionic model $\Op{\omega}_j= \Op{c}_j$ \eref{clif},
$\OpS=\OpS_R \in \mathrm{Spin}(2d)$ and $S_{jk}$ corresponds to orthogonal 
matrix $R$ \eref{SpinR} and conserving number of fermions is
taken into account in \eref{USpin} with unitary matrix. 

Therefore, for conserving number of boson and fermions evolution can be described
by the same unitary group with ``reduced'' dimension $d$ instead of $2^d$ for
most general quantum circuit and significant difference in complexity of 
amplitudes evaluation for multi-mode measurements looks especially challenging.

The fermionic model may be effectively simulated by classical computer producing
some controversy with bosonic case.
On the other hand average photon numbers in output of each mode can be 
simply calculated likewise with fermionic case. 
So, classical computer could effectively simulate output for each mode, 
but without proper quantum multi-photon correlations. Recent
achievements in experiments with photons \cite{lop14,ulo15} permit
to check such subtleties.

\smallskip

For {\em abstract bosons model} introduced in \sref{Sec:Naive} unitary 
operator $\OpS$ acting on states
could be compared with $\Op{u} \in \mathrm{SU}(d)$ in \eref{upsi}, but without analogues of operators such as 
$\Op{\omega}_j$ in \eref{GenTrans}. 

A Fock state \eref{FockSt} may be considered as an analogue of
\eref{symnorm} in {\em abstract bosons model}, but confusing them may produce
certain problems.
Let us consider simplest case with action of $\OpU_U$ on a basis of $\Hil$
\begin{eqnarray}
\fl
\ket{k} = \Op{a}_k^\dag\vac, \quad
 \OpU_U \ket{k} = \OpU_U\Op{a}_k^\dag\vac = \OpU_U\Op{a}_k^\dag\OpU_U^{-1}\OpU_U\vac \nonumber\\
\quad =\sum_j U^*_{jk}\Op{a}_j^\dag\theta_{\OpU}\vac 
 =\theta_{\OpU}\sum_j U^*_{jk}\ket{k} 
 =\theta_{\OpU}U^*\ket{k},
\label{UFock} 
\end{eqnarray}
where $|\theta_{\OpU}|=1$ was already introduced earlier, $\vac$ denotes vacuum state,
and $\ket{k}$, $k=1,\ldots,d$ correspond to basis of $\Hil$. In such a case in
{\em abstract bosons model} difference between $\OpU_U$ and $U^*$ 
lacks of proper treatment and simply missed sometimes. 
On the over hand \eref{USum} and \eref{MUU}
\begin{equation}
  \Op{a}'_j = \sum_{k=1}^d U_{jk} \Op{a}_k
  = \OpU_U \Op{a}_j \OpU^{-1}_U.
  \label{UUU}  
\end{equation}
would not make a sense, if $\OpU_U$ and $U^*$ are not distinguished.
Indeed, both signs $\pm\,\OpU$ lead to the same matrix $U$ in \eref{UUU}
and define of $\OpU_U$ as an operator from {\em double cover} 
of unitary group.

\smallskip
 
Together with the necessity in consistent mathematical expressions 
there are also physical
reasons to make distinction between two models discussed above. The oscillator 
model based on quantization of classical linear optics
and the {\em abstract bosons model} is close related with quantum
approach to discrete models such as group of permutations \cite{Weyl}.
In such a way the subtle relations between two models may be 
compared with a wave-particle duality.

\end{document}